\documentclass[runningheads]{llncs}
\usepackage{amsmath}
\usepackage{amsfonts}
\usepackage{graphicx}
\usepackage{algorithm}
\usepackage{algorithmicx}
\usepackage{algpseudocode}
\usepackage{cleveref}

\begin{document}
\authorrunning{Konstantin Sorokin, Andrey Zaitsew et al.}
\title{Global cognitive graph properties dynamics of hippocampal formation
}
%
%
\author{Konstantin Sorokin\inst{1} \and
Andrey Zaitsew\inst{2} \and
Aleksandr Levin\inst{1}  \and German Magai \inst{1} \and Maxim Beketov \inst{1} \and Vladimir Sotskov \inst{3}}
%
\institute{Higher School of Economics,
International Laboratory of Algebraic Topology and Its Applications
\email{ksorokin@hse.ru}\\
\and
Higher School of Economics, Faculity of Computer Sciences
\and
Laboratory of Neuronal Intelligence
Institute for Advanced Brain Studies Lomonosov Moscow State University}
\maketitle              
\begin{abstract}
In the present study we have used a set of methods and metrics to build a graph of relative neural connections in a hippocampus of a rodent. A set of graphs was built on top of time-sequenced data and analyzed in terms of dynamics of a connection genesis. The analysis has shown that during the process of a rodent exploring a novel environment, the relations between neurons constantly change which indicates that globally memory is constantly updated even for known areas of space. Even if some neurons gain cognitive specialization, the global network though remains relatively stable. Additionally we suggest a set of methods for building a graph of cognitive neural network.

\end{abstract}
\section{Introduction}

The idea of building a graph of connections in terms of contextual connectedness are inspired by ideas of functional connectome and cognitome of K.V. Anokhin \cite{KV}. The initial dataset we are working with are neuronal activities of CA1 region of hippocampal formation registered using calcium imaging. Then using different correlation dynamics of the neuronal dynamics time series the graph is constructed and natural graph properties are studied. We intentionally didn’t focus on neurons specializations and checked if the properties of global networks would change while an animal is performing a learning of new environment. We have already shown in the previous study that neurons (place cells to be particular) have quite rapid tuning dynamics \cite{ours} in terms of specialization. Current study is aimed to answer the question: does it affect the global properties of a neural network, or, say, cognitive map? Thus question comes naturally from the referenced study as it is shown there that some neurons, which are attached to certain places instantly gain their specialization, whereas some need several more visits of a certain place on an arena to stabilize. 
We have also previously showed that cognitive map is very useful in tasks of topology reconstruction of the ambient environment using place cells activities only \cite{ours2}. In this study we also used a time series of preselected place cells as one of methods of constructing the covering of a arena, where an animal was freely navigating. This method showed to be meaningful but global network forming dynamics is still yet to be learned. 
Studying the graph of neural network contextual coactivation connections would help understanding the process of learning (in general) and exploring novel environment of arena by a rodent. Looking at such graphs in dynamics may be extremely useful for describing the way how rodent understands and learns the space around as well as how the space memory is translated into the synapses and synaptic strengths. 

Having in mind all these results, we can discuss the way the network would perform, in general, on a cells level. It is clear that it has to retain some stability of it's properties, by design, and yet it should be efficient from the signal translation point of view. But how does it change when the network performs active reconstruction while animal is learning something new (we have shown \cite{ours2} that it in fact does)? The efficiency of a network can be shown with the small-worldness \cite{base}, for example which is characterized by certain properties, like the clusterization coefficient and average path legnth. 

While neural networks which used in machine learning are single or bidirectional, the biological neural networks may be multidirectional. The STDP (Spike-timing dependent plasticity and the cognitive map) \cite{stdp} \cite{stdp2} suggests that if there are two neurons connected with a synapse and the presynaptic is fired after post-synaptic one, the synapse strength reduced and if vice-a-versa the synapse-strength is reduced. In computer science words, the brain tries to greedely reduce path length.

Another important metric showing the effectiveness of the network is presence of rich-clubs \cite{rich}. Rich-clubs are observed in many biological systems, but for cognitome they are especially reasonable as a system of distributing information between different subsystems.
Such rich clubs may also migrate, i.e. change the neural connections, though we would expect our graph to have at least one subgraph with very rich collection of connections.

Additionally we expect that most of neurons will have synapses as the neural network should connected. In case if the connection amount is too small this means that actually neural network unable to store and transmit information. Moreover, these networks wouldn't perform robustly in case of reconstruction of unexpected damage of some parts. 
Such networks would not give a significant results which would be useful in further research.

\section{The experiment and data preparation}
\subsection{Experimental setup}
During the experiment, four different mice (C57Bl/6J) were each placed in a cylindrical arena with obstacles. A single-photon detector was installed on each mouse's skull to record neuronal activity in the CA1 region of the hippocampus. An ion-dependent dye was used for detection, which fluoresces when interacting with calcium ions emitted in synapses. More detailed information on the surgical interruptions performed on animals can be found in our previous paper \cite{ours2}. 

The experiment was conducted over three days with the same mouse, and each day a different number of obstacles was placed in the arena. So, for each mouse three experiments were performed. While the mice explored the arena, their activities were recorded on a vertically oriented camera. The video with the movements of the animals was processed, adjusting the brightness and contrast to distinguish the mouse from other objects falling into the frame. Two colored LEDs were attached to each mouse's head to restore the direction of the animal's head.

The wire coming from the camera miniscope attached to the mouse's head was deliberately made white. Animal's movements were synchronized with the readings of the neural activity sensor and analyzed.


\subsection{Highlighting of neurons}
\begin{figure}[!ht]
 \centerline{\includegraphics[width=1\textwidth]{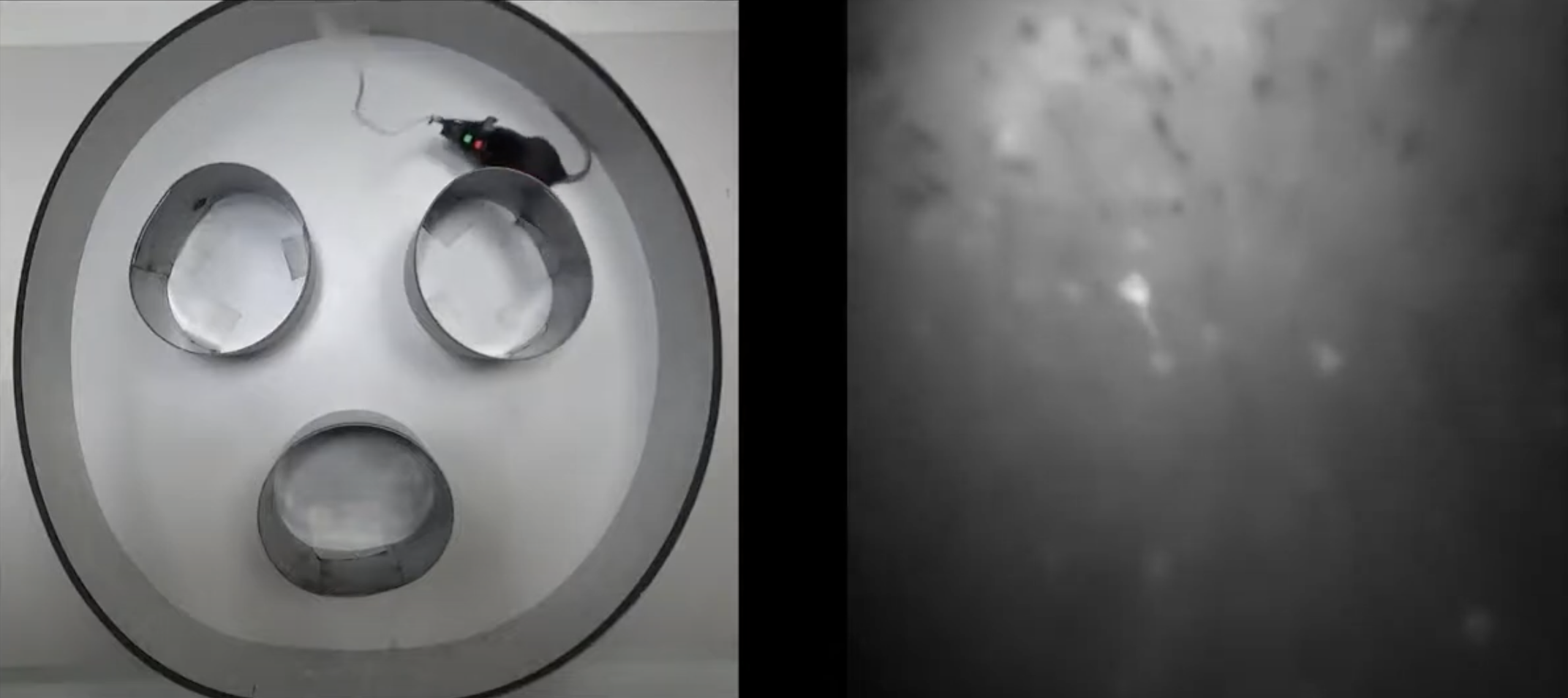}}
 \caption{Mouse and data from the miniscope attached to it}
 \label{mouse}
\end{figure}

Data from the camera is a black and white video recorded in 20 frames per second (see \cref{mouse}). There are multiple problems related to this: firstly, due to mechanical influences on the miniscope, the use of these recordings is impossible without post-processing like stabilization and the overall motion correction; secondly, this is not really a computer-friendly data format. In this work the CaImAn package \cite{caiman} is used to solve these problems. 

For the motion correction NoRMCore is used. This algorithm divides the area into intersecting sections, which are processed separately, and are then combined using smooth interpolation.

After the stabilization the neurons are identified with CNMF-E, a one-photon extension of the CNMF framework, which stands for Constrained Nonnegative Matrix Factorization. 

After that we obtain a smooth signal of each candidate to become a neuron, which need more processing. We used several criteria to filter out the ''fake'' neurons which do not glow uniformly, have inconsistent activation patterns, etc. At the following step the chosen neurons are normalized and uniformed. Finally, we have a set of time series for each neuron represented as $a = \{ a_1, a_2, \dots, a_t | 0 \leq a_i \geq 1 \} $ and  $i \in I $ which is set of all frames of video.






\section{Graph constructions and similarity metrics}

The idea behind graph ($G = (V, E)$) construction is very simple and it is mostly about the matter of measuring similarity distance between vertices $V = \{ 0, 1, \dots, n \}$ of a graph, which are exactly the neurons with their activity. The edges $E$ then are directed and weighted and should be represented by a triplet $ (a, b , \hat{\rho} ) \in E $, where $a,b \in V$ and . Such triplet would contain number of source vertex, number of destination vertex and the weight. From neurobiological perspective it contains presynaptic neuron, post-synaptic neuron and the estimate of synaptic strength. In current definition graph is full, i.e. contains all possible edges. Such graph is unlikely to be useful in a research so we limit it's connections by the lower bound threshold $\underline{\rho}$ for the edge existence. then the set of edges is defines as $E = \{ (a, b, \rho) | \rho(a,b) > \underline{\rho} \}$.

However, we are interested in network dynamics, that's why we would slice the time series.
As previously mentioned we would like to design metric which can assess the influence of one’s neuron activity on another one. In order to do this we would need to compare the “previous” value of activity of presynaptic neuron
candidate with activity on a post-synaptic candidate neuron. So we would need to shift our vectors by $k$ entries($k$ should be adjusted with according to
the discretization rate, in our case $k = 1$ unless stated otherwise). The shift operation is defined as follows: $s(\overrightarrow{U}, k) = (U_1,U_2, \dots ,U_{T-k})$, where $k \in \mathbb{Z}_{>0}$, $T = {\dim}(\overrightarrow{U})$. For $k \in \mathbb{Z}_{<0}$: $s(\overrightarrow{U}, k) = (U_{k+1},U_{k+2}, \dots ,U_T)$. Intuitively saying the value $i$ if $s(\overrightarrow{U}
, k)$ would correspond to value $i + k$ in
$s(\overrightarrow{u},-k)$ for any $k \in \mathbb{Z}, k > 0$.

So, combining it into a value to measure the synaptic strength we define a function which maps two vectors of neural activity to some real number preserving
synaptic strength. Also we should remember that out vectors needs to be shifted as the effect of presynaptic neuron on the post-synaptic neuron is slightly delayed. The function is defined as follows:
$ \rho \colon \mathbb{R}^{T-k} \times \mathbb{R}^{T-k} \rightarrow \mathbb{R} $, where $T$ is the duration of the recording, and $k$ is fixed number by which the time series is shifted (Intuitively saying this our expectation on delay between the effect of presynaptic neuron on post-synaptic one). 

In order to build the graph we have used the following algorithm for given $\rho, \underline{\rho}, k, N$, where $N$ is a matrix of all neural activities

\begin{algorithm}
	\caption{Graph construction}\label{euclid}
	\begin{algorithmic}
		\Function{BUILD GRAPH}{$\rho, \underline{\rho}, k, N$}
		
		\State $h \gets \dim(N^{1})$ \Comment{Obtain a height of matrix}
		\State $w \gets \dim(N_{1})$ \Comment{Obtain a width of matrix}
		\State $V \gets  \{ 1, \dots , w \} $
		\State $E \gets  \{  \} $
		\For {$i \in \{ 1, \dots, h \}$} 
		\For {$j \in \{ 1, \dots, h \}, i \neq j$}
		\State $\tilde{\rho} \gets \rho((s(N^{(i)},k)), s(N^{(j)},-k)) $
		\If {$ \tilde{rho}) \geq \underline{\rho}$}
		\State $E \gets E \bigcup \{ (i, j, \tilde{\rho}) \}$.
		\State $i \gets i-1$.
		\EndIf
		\EndFor
		\EndFor
		\State $\textit{Result} \gets (V,E) $
		\EndFunction
	\end{algorithmic}
\end{algorithm}

For the function it's only a matter of defining the $\rho$, which we have several candidates for. 

\subsection{Correlation}

The correlation is the most common, obvious and simple way to measure effect of one variable to another. The correlation coefficient is a statistical measure that describes the degree to which two variables are linearly related. In context of our research, the correlation between the firing rates of two neurons over time
can be an efficient as a proxy variable for synaptic strength. For two vectors of neural activity $\overrightarrow{a}, \overrightarrow{b}$ the Pearson correlation measure would be as follows: $$ \rho(\overrightarrow{a}, \overrightarrow{b}) = \dfrac{\sum^n_{i=1} (\overrightarrow{a}_i - \overline{a})(\overrightarrow{b}_i - \overline{b})}{ \sqrt{\sum^n_{i=1} (\overrightarrow{a}_i - \overline{b})^2 \sum^n_{i=1} (\overrightarrow{b}_i - \overline{b})^2  }}, $$

where $ \overline{a} = \dfrac{\sum^{T-k}_{i=1} a_i}{T-k} $ and $ \overline{b} = \dfrac{\sum^{T-k}_{i=1} b_i}{T-k} $

\subsection{Granger causality}
The Granger causation \cite{gr} is a way to measure the effect of one time series on another by building a predictive models for each of time series. It was initially developed for econometric applications, where the signals may not have obvious and immediate interactions between each other and as the casualty is very important in neurons interaction we decided to test it. Granger causality may be useful as its already implies directionality of the interaction, i.e. we do not need to shift time series. According to the Granger causality test, if a signal $X$ “Granger-causes” (or “$G$-causes”) a signal $Y$, then past values of $X$ should contain information, which helps predict $Y$ above and beyond the information contained in past values of $Y$ alone.

The idea of Granger causation lies in fitting two autoregressive models and if $y(t)$ prediction is significantly improved by adding $x(t)$ to the model then $x(t)$ $G$-causes $y(t)$. The models itself is as follows:  $ y(t) = \sum^n_{i=1} \alpha_i y(t-i) + \epsilon(t) $, $ y(t) = \sum^n_{i=1} \alpha_i y(t-i) + \sum^n_{i=1} \beta_i y(t-i) + \epsilon'(t) $. 

Here, $\alpha_i$ and $\beta_i$ are the model parameters, $n$ is the model order, and $\epsilon(t)$ and $\epsilon'(t)$ are error terms. If the second model significantly improves the prediction of $y(t)$ over the first model, then we conclude that $x(t)$ Granger-causes $y(t)$.  Obviously, we can consider $x(t) = \overrightarrow{a}$ and $y(t) = \overrightarrow{b}$ 

Though the significance as is can not be used to measure and compare synaptic strength between neurons, as we actually would like to see only edges which are results of actual presence of synapses we may use a $p$-value of a test as a measure of synaptic strength along with setting the $p$ to 0.95 which corresponds to the $95\%$ level of statistical confidence. Additionally we set shift to $k = 0$ in this approach due to the autoregressive nature of models used in Granger casualty it already takes in account all required shifts of data.

\subsection{Coherence}

Coherence is another measure of the linear correlation between two signals. The coherence $C_{xy}(f)$ between two signals $x(t)$ and $y(t)$ is defined as follows: $$ C_{xy} (f) = \dfrac{|G_{xy} (f)|^2}{G_{xx}(f) G_{yy}(f)}, $$
where $G_{xy} (f)$ is cross-spectral density of $x(t)$ and $y(t)$, $G_{xx} (f)$ and $G_{yy} (f)$ are the power spectral densities of $x(t)$ and $y(t)$, respectively. The
cross-spectral density is the Fourier transform of the cross-correlation function, and the power spectral density is the Fourier transform of the autocorrelation
function, so $f$ here denotes frequency. The cross-spectral density is defined as follows: $ G_{xy} (f) = F(R_{xy} (\tau))$, where $F$ denotes the Fourier transform, and $R_{xy}(\tau)$ is the cross-correlation function of $x(t)$ and $y(t)$. The coherence also ranges from 0 (no coherence) to 1 (perfect coherence). 

In the context of the neural activity analysis the coherence would show the consistency in activation of different neurons on different frequencies. As synapse is working on different rate in the same manner we expect to observe a high coherence of the connected neurons on all of the rhythms(i.e. frequencies) in which hippocampus has worked. The synaptic strength does not depend on the frequency and this means that the connections having the higher average coherence indicate the most powerful synaptic strength. In spite of this we decided to use the mean of coherences calculated on frequencies from 1Hz to 100Hz. The following formula was used in order to calculate coherence between two neurons is as follows: $$ \rho(\overrightarrow{a}, \overrightarrow{b}) = \dfrac{1}{\overline{f} - \underline{f}} \sum^{\overline{f}}_{i = \underline{f}} C^D (\overrightarrow{a}, \overrightarrow{b}),  $$
where $C^D$ is discretized version of $C_{xy}, \underline{f} = 1, \overline{f} = 100 $.

\subsection{Performance comparison}

We have used dataset recorded when rodent was on 3-obstacles (which is a maximal number we used across experiments) arena for evaluation of the methods suitable for constructing graph of neural activities. Indeed, conjecturally, it's the most complex environment for an animal to explore and memorize. 

Even on preliminary part of testing classical Pearson correlation showed very high level of connectivity, but at the same time the network had an issue of multi-edges. We assumed that the reason for excessive multi-edges could be too low value of $\underline{\rho}$. Moreover, as the graph is not simply-connected (which is perfectly fine) even the small graphs were with multi-edges. In order to verify that the reason for multi-edges is not low value of $\underline{\rho}$ we have also tried use higher correlation boundary. The graph had had much less connectivity between edges, but multi-edges persisted. This indicates that correlation can not be used as metrics for building graphs with use of this approach as it can not be used for distinction of post-synaptic and presynaptic neuron creating multi-edges which leads to loss of the information on data directionality in neural network. 

\begin{figure}[!ht]
 \centerline{\includegraphics[width=1\textwidth]{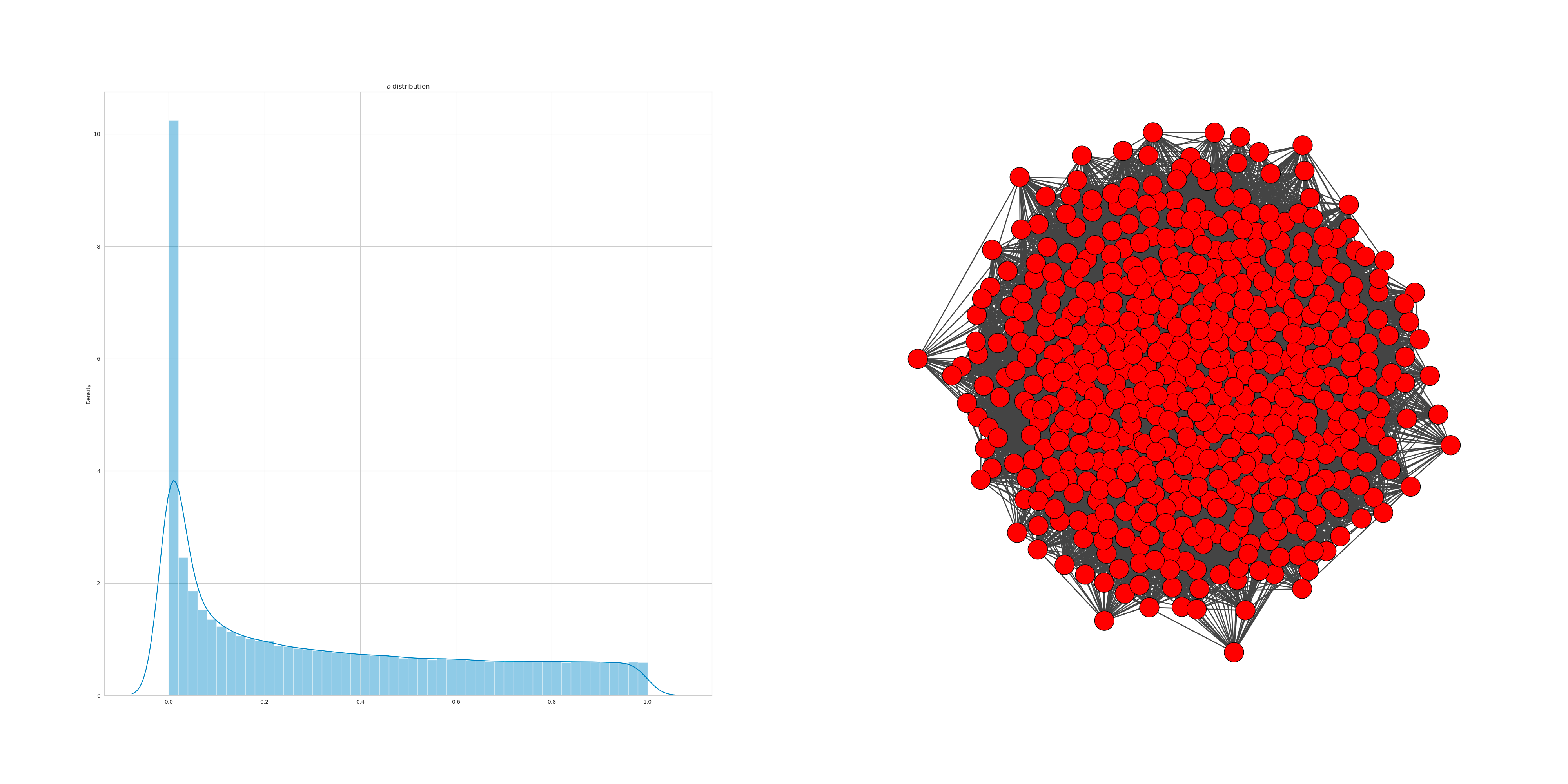}}
 \caption{Distribution of $\rho$ for Granger causality (on the left) and the graph built with this method with $\underline{\rho} = 0.95$ (on the right)}
 \label{granger}
\end{figure}

The results of Granger-casuality also gave poor results. In fact, it performed the worst results among the methods suggested. Taken the values of $\rho$ for Granger causation at \cref{granger} (left) shows that after 0.3 the distribution becomes almost flat. This is likely means that the graph would have too much edges. This is confirmed when the graph is built \cref{granger} (on the right).

The likely reason for this is stability of a series and noise dependency of observed values of neural activity. This leads to that the autoregressive model predict good just having one previous value of neural activity and may improve prediction being supplied other neuron data. The improvement of such method may be filtering out only spikes making activity completely binary. Though such would also lead to the too good predictions of autoregressive models as most neurons do not express spike activity often enough.

\begin{figure}[!ht]
 \centerline{\includegraphics[width=1\textwidth]{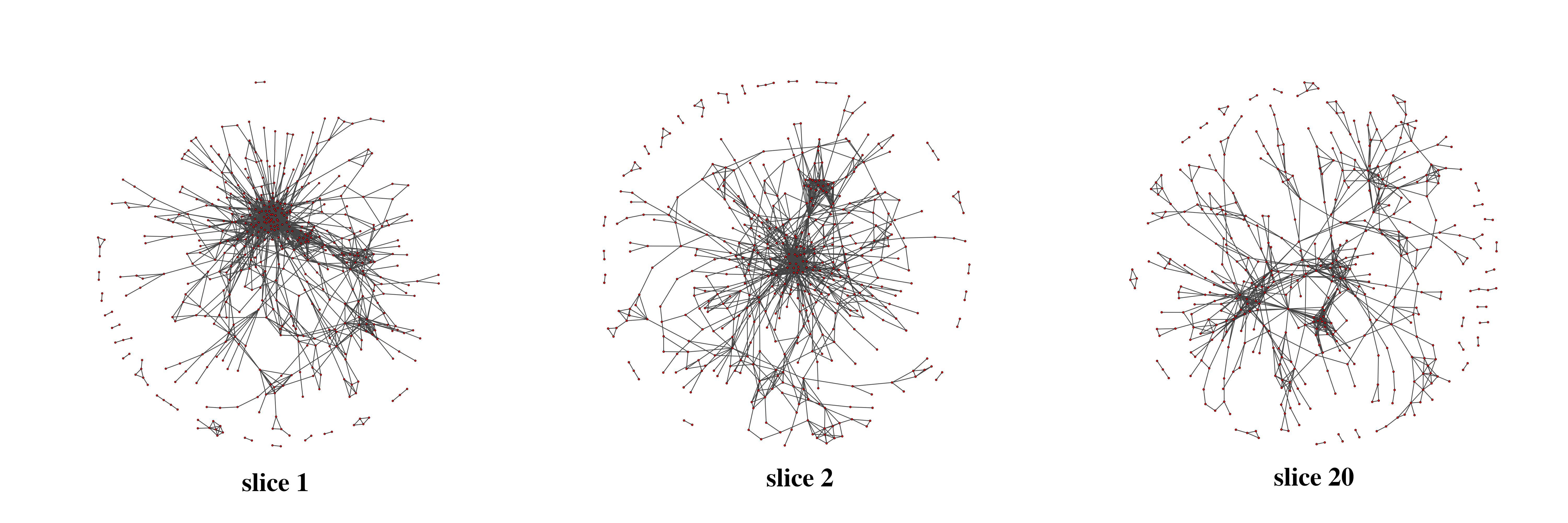}}
 \caption{Coherence based graph on time-partitioned data}
 \label{coherence}
\end{figure}

The coherence being theoretically the most suiting solved the problems of the previous approaches. It does not give multi-edges and has very stable graph connections across slices \cref{coherence}. The computationally check for multi-edges shown the relatively small amount of them(about $0.5\%$ from overall edges generated on all slices had complementary counter edge). Also, the significant part of neurons are included into the path-connected network and the clusters of the local ''rich-club'' are permanently formed. This all is in accordance to what was expected for the contextual network design.

Even from the several slices we checked during the coherence evaluation we can see the stability of the network properties, at the same time of high plasticity of connections. That's why in the following part we will study the global properties more precisely. 

\section{Dynamics}

\begin{figure}[!ht]
 \centerline{\includegraphics[width=0.7\textwidth]{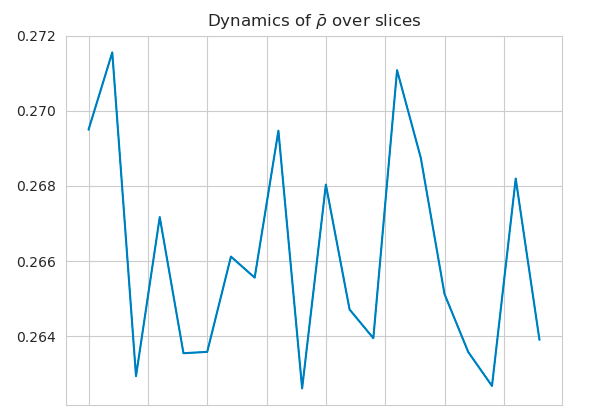}}
 \caption{Distributions of $\overline{\rho}_i$ of neural dynamics across slices of a single mouse}
 \label{dynamics}
\end{figure}

On \cref{dynamics} the average coherence ''metric'' values is shown in dynamics. Each point on this plot is given for a graph computed for $S_i$ with ‘BuildGraph‘ algorithm we described above, namely $G_i = (V,E_i)$ and calculated as follows: $\overline{\rho}_i = \dfrac{1}{|V|^2 - V} \sum_{i \in V} \sum_{j \in V, j \neq i} \rho(i,j)$. 

We may see from the graph \cref{dynamics} that $\overline{\rho}$ oscillates between 0.26 and 0.27 which indicates that on average synaptic strength has no trend and does not tend increase or decrease. Such behaviour along with the graphs presented earlier may indicate that rodent’s hippocampus tend to distribute the neural network.

\begin{figure}[!ht]
 \centerline{\includegraphics[width=1\textwidth]{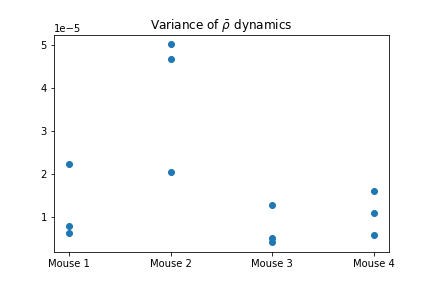}}
 \caption{Distributions of variance of neural dynamics across slices across all experiments with all mice}
 \label{var_rho}
\end{figure}

We applied this approach to all the experiments we had. So we had distributions for all four mice (see \cref{var_rho}), by 3 experiments on arenas with 1, 2 and 3 ''holes'', so the environments explore were not the same. Th distribution of variance of $\overline{\rho}_i$ for each mouse shows strong stability across experiments with the outliers still being very minor in value. 

\begin{figure}[!ht]
 \centerline{\includegraphics[width=1\textwidth]{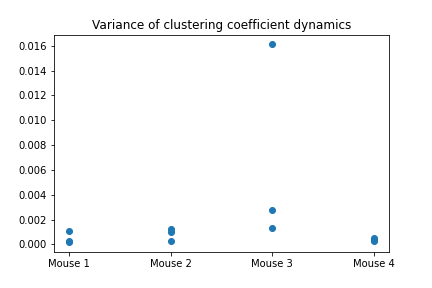}}
 \caption{Distributions of variance of graphs clustering dynamics across slices across all experiments with all mice}
 \label{var_cl}
\end{figure}

We also decided to check the network dynamics property by calculating the clustering coefficient for vertex $u$ in $G_i$ (a graph of certain slice of time series) as follows: $$c_{u,i} = \dfrac{1}{\deg(u) (\deg(u)-1 )} \sum_{v,w \in V} (\widetilde{W}_{u,v}, \widetilde{W}_{u,w}, \widetilde{W}_{v,w} )^{1/3}$$, where $\deg(u)$ is a vertex degree in corresponding graph, $\widetilde{W}_{u,v} = \dfrac{W_{u,v}}{\max(W_{u,v})}$  is normalized weight of edge between $u,v$ and equals 0 if there's no edge between them. The clustering coefficient for all graph is the averaged clustering coefficient of all vertices: $c_i = \dfrac{1}{V} \sum_{u \in V} c_{u,i}$. The variance of clustering coefficients for all experiments is shown at \cref{var_cl}. 

It varies in a wider range, obviously, but still it is clearly very stable with only one outlier out of 12 experiments. In general clustering coefficient trends and values demonstrate that the network is constantly changing, which means that even when mouse learns the arena completely, the network in general recombines in other areas, not related to place cells. It might be related to other mice activities. But the general network properties also on the clustering coefficient level remain within a very certain limits.

\section{Conclusion}

In this work we have shown that the coherence is very promising method for constructing the neural network in dynamics. It performs gradual reconstructions of a network and gives a set of graphs usable for studying. The network analysis showed that it's properties, even during reconstruction of local parts when learning the environment are very stable and remain within tight limits. It shows the robustness of a hippocampal neural network and gives a new perspective to future modeling of such networks. As further research we may suggest the following: analyse graphs in more depth, develop models for activity and develop model of synaptic strength change. The further analysis may be focused on developing of more quantitative methods of assessing resulting graphs. Additionally the analysis can include different parameters of a network like graph hyberbolicity, small-worldness and hierarchical properties.

\section{Acknowledgements}
The article was prepared within the framework of the project “Mirror Laboratories” HSE University, RF.
This research was supported in part through computational resources of HPC facilities at HSE University

\section{Ethics statement} The animal study was reviewed and approved by all methods for animal care and all experimental protocols were approved by the National Research Center “Kurchatov Institute” Committee on Animal
Care (NG-1/109PR of 13 February 2020) and were in accordance with the Russian Federation Order Requirements N 267 M3. Three C57BL/6J mice were used in this study, ages ~2–3 months. Mice were used without regard to gender.

\end{document}